\documentclass[floatfix,reprint,pre]{revtex4-2}
\usepackage{graphicx} %
\usepackage{amsmath,amsfonts,amsthm,amssymb,bm}
\usepackage{float}
\usepackage{color}
\usepackage{algorithm}
\usepackage{algpseudocode}
\usepackage{appendix}
\usepackage{booktabs}
\usepackage{url,hyperref}

\begin{document}
\title{Clock-state olfactory search in turbulent flows using Q-learning: The geometry of plume recovery}
\author{Marco Rando}
\affiliation{Universit\'e C\^ote d'Azur, Inria, CNRS, Laboratoire J.A. Dieudonn\'e, 28 avenue Valrose, 06108, Nice, France}
\author{Robin A.\ Heinonen}
\affiliation{Machine Learning Genoa Center \& Department of Civil, Chemical and Environmental Engineering, University of Genova Via Montallegro 1, 16145 Genoa, Italy}
\author{Yujia Qi}
\affiliation{Machine Learning Genoa Center \& Department of Civil, Chemical and Environmental Engineering, University of Genova, Via Montallegro 1, 16145 Genoa, Italy, now at University of California, Dept Mechanical Engineering, Engineering II, Santa Barbara, CA 93106-5070, USA}
\author{Agnese Seminara}
\affiliation{Machine Learning Genoa Center \& Department of Civil, Chemical and Environmental Engineering, University of Genova, Via Montallegro 1, 16145 Genoa, Italy}
\date{\today}

\begin{abstract}
    Finding an odor source in a turbulent flow requires effectively leveraging the history of olfactory observations into a robust navigation strategy. In this work, we use tabular Q-learning to train an olfactory search agent with a minimal memory of past observations: only a running clock since the last whiff. This agent learns an interpretable strategy to recover the plume which combines well-known behaviors observed in insects: surging, casting, and a return downwind. While achieving good performance on data from direct numerical simulations of turbulence, the agent is limited by an inability to adapt its strategy to the local intermittency level; we show that providing more flexibility improves robustness.
\end{abstract}
\maketitle
\section{Introduction}
Insects frequently track sources of odors to locate mates and food. In the well-studied and relevant setting of a turbulent flow (such as the atmospheric boundary layer), this ``olfactory search'' behavior can be remarkably complex, having evolved to cope with the steep challenges posed by stochastic odor encounters, odor sparsity, and the absence of useful spatial gradients, all of which are endemic to turbulence \cite{carde2008,reddy2022,steele2023}. It is widely recognized that recovering the plume is key to efficient navigation both in biology and robotics, as agents often encounter extended odor-free regions in turbulent conditions, e.g.~\cite{carde2021_annurev,celani2014,heinonen2025optimal,rando2025,ishida2006mobile}.

Evolved search strategies bear certain similarities across species. After detection of an attractive odor cue, insects frequently begin flying or walking upwind for several seconds~\cite{david1983,alvarez-salvado2018,kuenen1994,vanbreugel2014,demir2020}. If after this brief surge they do not detect odor, they typically engage in a variety of exploratory behaviors to recover the plume, for example looping \cite{vickers1996}, crosswind zigzags (possibly performed while drifting upwind or downwind) \cite{baker1987,mafra1996}, or direct downwind motion \cite{willis1991,kuenen1994}. The precise sequence of behaviors depends on the species; for instance, casting is highly stereotyped in flying moths but can occur with variable timing, speed, and structure in flying \emph{Drosophila} and walking insects such as cockroaches \cite{steele2023}. 

The properties of the plume and the turbulent flow help shape these recovery behaviors, which have been observed to vary when the plume structure is changed \cite{soman2026}. For example, from a Bayesian perspective, the cast-and-surge behavior may be understood as a means to exhaustively explore the support of the prior on the source location, itself induced by the plume shape \cite{balkovsky2002,heinonen2025optimal,rigolli2022}. 

Recovery behavior is also influenced and constrained by the structure of the searcher's memory. Memory is necessary both to chain distinct behaviors together and to accumulate information about the source, which odor cues grant only noisily due to turbulence. More generally, memory is essential to any partially observable decision task \cite{kaelbling1998}. Behavioral and neural studies suggest that insects integrate information from odor detected along their path~\cite{alvarez-salvado2018,demir2020,jayaram2023,kathman2025}. The first relay of olfactory processing shows that insects are able to measure complex features of odor time traces \cite{brandao2021,patel2021}, but the extent to which these complex features are used for navigation is unknown. 

Several models of the agent's internal memory state have been explored in previous work, ranging from complex, high-dimensional representations such as a Bayesian posterior \cite{vergassola2007,loisy2022,heinonen2023} or the hidden state of a recurrent neural network \cite{singh2023}, to simple discrete representations, such as a finite-time window of the recent odor observation history \cite{rando2025}, a small handful of learned, abstract memory states (each encoding a distinct behavior) \cite{verano2023}, or a clock that resets at each odor detection \cite{balkovsky2002}. This last memory structure may be motivated by experiments on walking flies which suggest that the time since odor was last detected is an important feature driving behavior \cite{demir2020,jayaram2023}.

In this work, we ask what shapes the recovery strategy of efficient turbulent navigation. To this end, we develop an algorithm that learns how to navigate a realistic turbulent odor plume, using only a clock state that encodes time since the last odor encounter, thereby isolating how much of the navigation problem can be captured by a minimal recovery-oriented memory. This generalizes previous work hard-coding a heuristic, bio-inspired cast-and-surge strategy \cite{balkovsky2002}; here, the recovery strategy is optimized using Q-learning. It is also a simplification of Ref.~\cite{rando2025}, which learned an additional in-plume strategy depending on the recent odor observation history. 
As in~\cite{rando2025}, we assume agents are aware of the direction of the mean wind, which allows them to learn policy of actions labeled as upwind, downwind and crosswind; the more realistic case where agents can only rely on instantaneous wind speeds is investigated elsewhere~\cite{piro2026}.

We find that remembering only the time since last detection is sufficient for efficient navigation, showing that a highly compressed memory can capture much of the performance of more elaborate strategies which include more information about the odor observation history. This agrees with previous work finding that recovery is by far the most important component of the search strategy \cite{heinonen2025optimal}.%

We show that a clock-state Q-learning agent learns a structured recovery policy consisting of upwind surge, lateral exploration, and delayed downwind return. 
Importantly, the details of the policy vary in different realizations of the training, confirming that a multitude of recovery strategies reach similar performance.
We find that the geometric characterization %
of these behaviors has %
a systematic dependence on plume sparsity. 
We argue that the main drawback of this agent is a lack of adaptability to its location within the plume, and we compare our results to a slightly richer two-mode extension which improves generalization to distinct plumes. 

\begin{figure}
 \includegraphics[width=\linewidth]{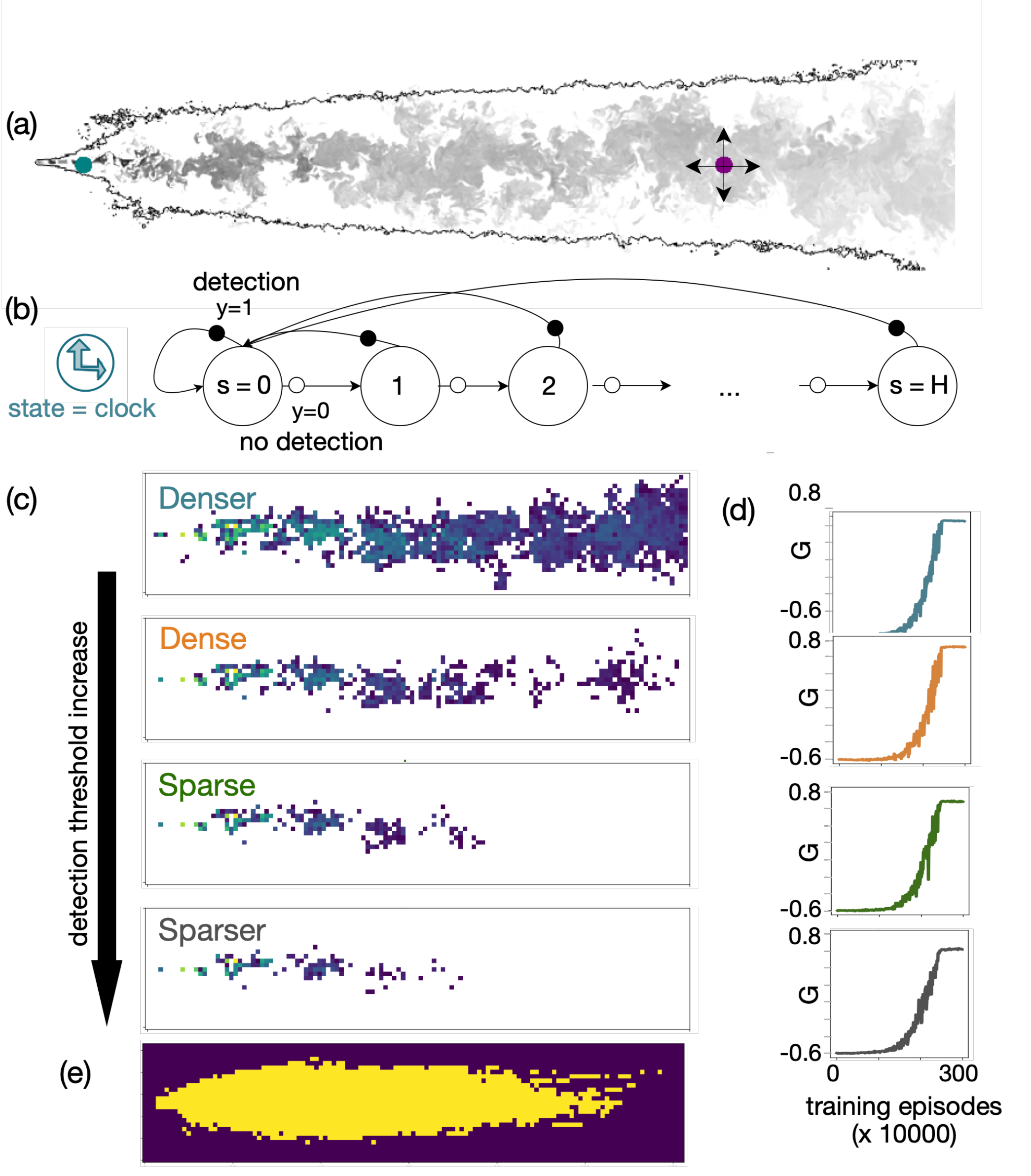}
 \caption{Problem setting. (a) We perform simulations of a turbulent flow with a source of odors (blue circle). An agent (purple circle) is sensitive to odors and can move in the four cardinal directions. (b) The agent's internal state $s$ is a clock measuring the time since the last odor detection, running from $s=0$ up to $s=H,$ where $H$ is the duration of an episode. Odor detections (black circles) reset the clock, while odor blanks (white circles) increment it. (c) We define four environments with different sparsity levels, which we set by varying the odor detection threshold. (d) Episodic reward $G$ as a function of number of training episodes. The Q-learning algorithm converges within $3\times 10^6$ episodes. (e) Mask showing possible initial positions of the agent within the plume. The position is drawn uniformly from this set, and the starting simulation snapshot is selected randomly.}\label{fig:fig1}
 \end{figure}

The remainder of this paper is organized as follows. We present our results in Sec.~\ref{sec:results}: first, we benchmark the performance of the Q-learning strategy agent; second, we define metrics characterizing the geometry of trajectories, and by studying their values in Q-agents, we show how sparsity shapes the recovery strategy; third, we introduce a more sophisticated agent with two independent recovery strategies, and show these strategies adapt to different regions of the plume; and fourth, we demonstrate that a quasi-optimal Bayesian agent can adapt its strategy depending on its observation history, a flexibility which the Q-learning approach lacks. We then summarize our results and suggest future avenues of research in Sec.~\ref{sec:discussion}. Finally, in Sec.~\ref{sec:methods} we conclude with a detailed presentation of our methodology.

\section{Results}\label{sec:results}

\begin{figure}
 \includegraphics[width=\linewidth]{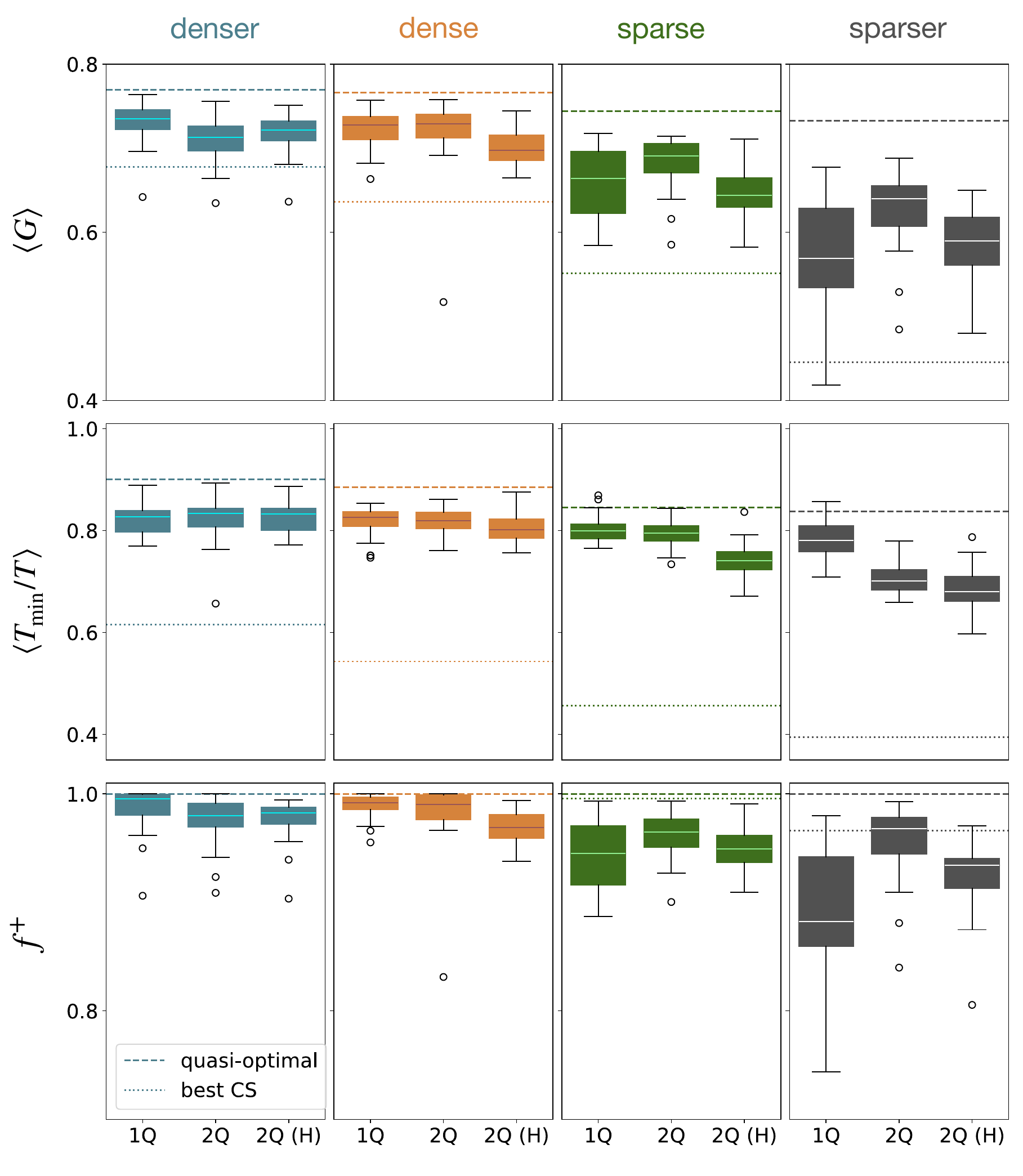}
 \caption{Performance of the agents presented in this paper. For each strategy, we optimize the mean episodic reward $\langle G \rangle$ (top); to interpret the meaning of $G$ we additionally quantify the mean normalized time $\langle T_{\min}/T\rangle$ (center) and the success rate $f^+$ (bottom). The single-Q, two-Q and heuristic two-Q strategies are shown as box-and-whisker plots over an ensemble of twenty different trained agents (outliers are indicated by white circles). Two baselines, a quasi-optimal Bayesian agent and an optimized one-parameter cast-and-surge heuristic, are presented as horizontal lines}\label{fig:fig2}
 \end{figure}

Reinforcement learning \cite{sutton1998} provides a natural framework for olfactory search and other problems involving decision-making under uncertainty. Q-learning \cite{watkins1992} is particularly easy to train and is adapted to scenarios where the agent's internal state space and action space are both discrete. To explore how recovery strategies adapt to the environment, we simplify the Q-learning approach developed in Ref.~\cite{rando2025}. In that work, we crafted a small set of olfactory states from a fixed temporal window of concentration measurements; in the event where all measurements in memory are zero, a separate recovery strategy was learned using the time since the last odor whiff as the state. Presently, the olfactory states are dropped, and the agent remembers only the time $s$ since the last odor ``whiff'', defined as odor being above a fixed concentration threshold. The agent therefore learns a fixed sequence of moves as a response to plume loss.

The methodology, presented in more detail in Sec.~\ref{sec:methods}, is illustrated in Fig.~\ref{fig:fig1}. We train and test our approach on high-quality direct numerical simulation of a turbulent channel flow, previously used in Refs.~\cite{rigolli2022,rando2025}. The flow has an embedded point source of contaminants and represents similar conditions to those of the lower atmospheric boundary layer. We test four different detection thresholds on the concentration, allowing us to tune the sparsity of encounters with the odor. For consistency, the agent's initial time and position in the simulation is drawn from the same distribution for each choice of threshold.

\subsection{Performance}
We first test the raw performance of the Q-learning algorithm. We measure this using the average episodic reward $G$, which is the quantity that is directly optimized by the algorithm.  
The result is plotted in Fig.~\ref{fig:fig2}; we compare the results to both a quasi-optimal Bayesian agent and a cast-and-surge heuristic whose characteristic opening angle has been tuned to the setting ~\cite{baker_cast_and_surge,balkovsky2002}. 
The Q-learning agent significantly outperforms cast-and-surge, highlighting the inefficiency of a hard-coded recovery strategy and the value of optimizing the timing and geometry of recovery behaviors. 
Note that the learned policy and its performance varied significantly from training run to training run; hence, in Fig.~\ref{fig:fig2} and elsewhere we will generally show box-and-whisker plots to illustrate the distribution of learned policies. The upper end of the whisker corresponds to the best learned policy. Somewhat surprisingly, the best Q learning policy nears performance of the Bayesian algorithm, while using a much simpler memory structure.\\
To rationalize the meaning of $G$, we represent additionally the normalized arrival time $T_{\rm min}/T$ (where $T_{\rm min}$ is the minimal time of arrival for the episode), and the success rate $f^+$. 
In all sparsity settings, the agent typically arrives to the source with a 90\% success rate or greater, approaching 100\% for the least sparse settings. 

\begin{figure}
 \includegraphics[width=\linewidth]{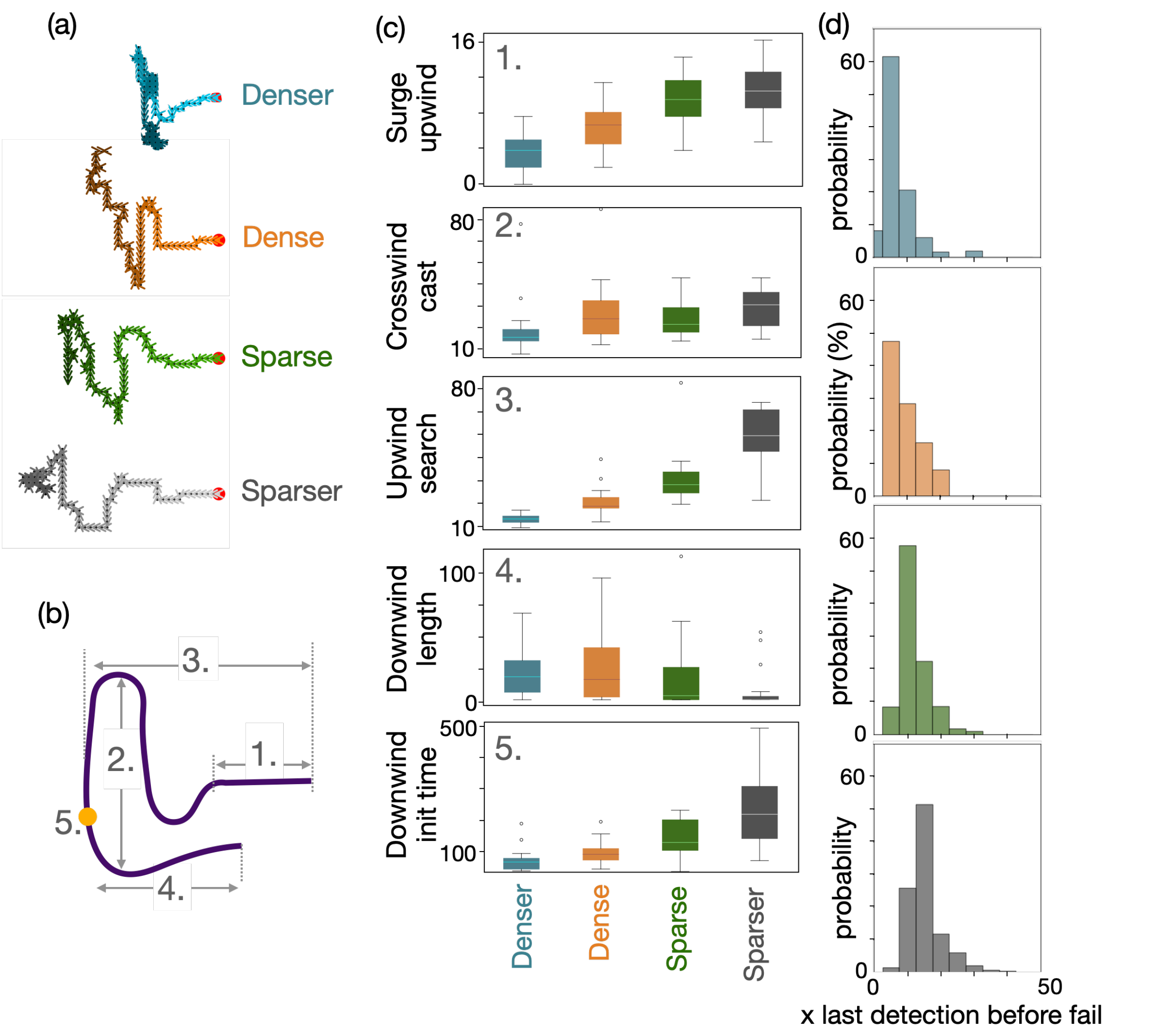}
 \caption{Geometry of the search strategy depends on sparsity of the plume. (a) Trajectories of (single-)Q agent in absence of odor for different plume sparsity levels, color coded as indicated.  Increasing values of the clock state are represented from bright colors ($s=0$) to dark colors ($s=H$). (b) Five different metrics are defined to characterize the shape of a recovery strategy: (1) the upwind surge length, (2) the crosswind cast width, (3) the total upwind search length before returning downwind, (4) the downwind return length, and (5) the downwind return initiation time. (c) Geometry of the learned policies in the different environments. The five metrics are shown as box-and-whisker plots across an ensemble of twenty trained agents. The plots show an increasing tendency to travel upwind and decreasing tendency to return downwind as plume sparsity increases. (d) Probability distribution of the downwind position of the agent at its last detection, conditioned on failing to find the target. The last detection occurs relatively close to the source in all environments, suggesting that failures occur when the agent overshoots the source. The peak moves downwind with increasing sparsity, indicating that agents get lost at larger downwind distances in sparser plumes. \label{fig:fig3}}
 \end{figure}

  \begin{figure}
 \includegraphics[width=\linewidth]{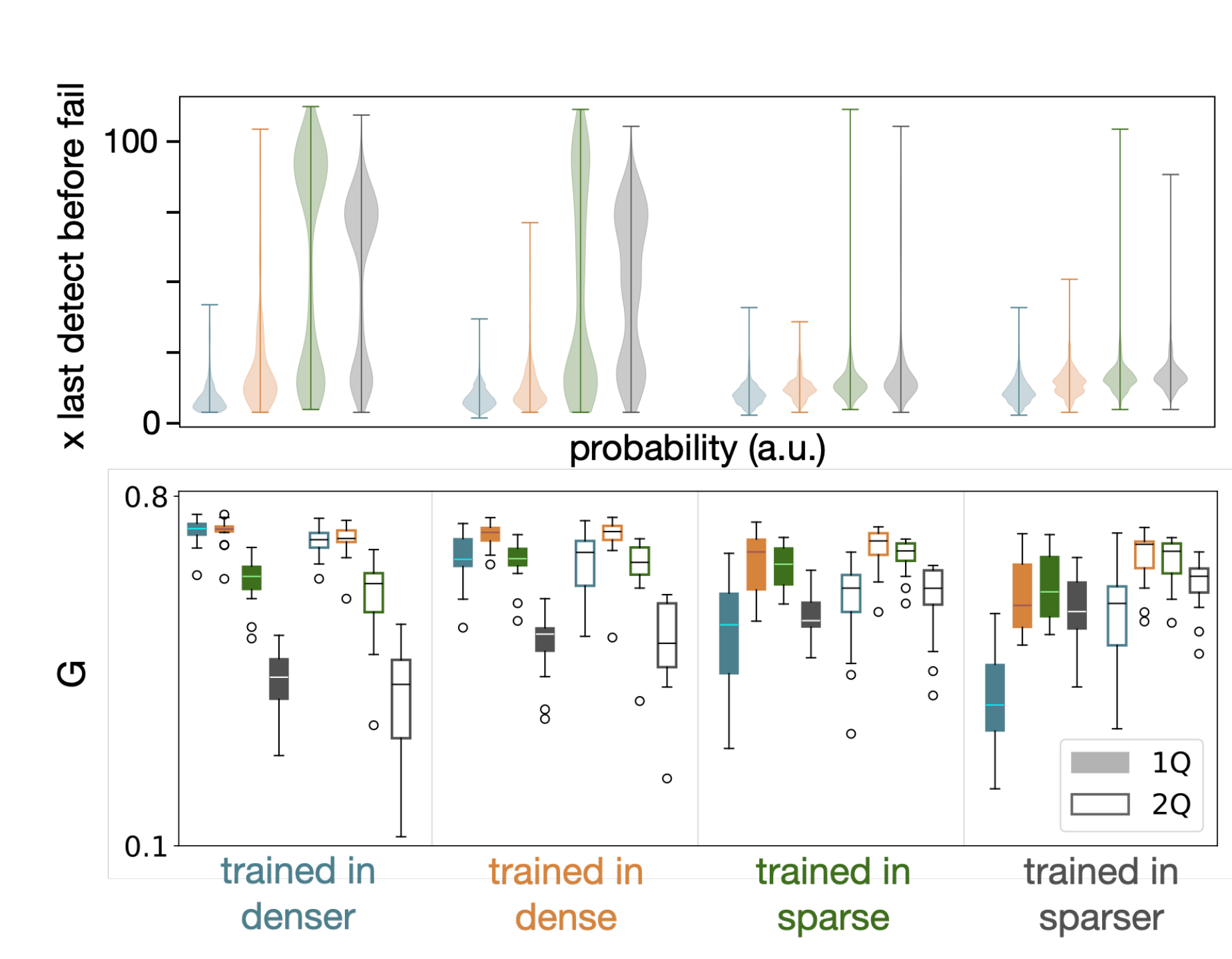}
 \caption{A Q-agent's generalization to different sparsity levels. Upper panel: violin plots showing distributions of the downwind positions of the last odor detection before failure, when an agent is trained in one environment (indicated in $x$-axis labels) and tested in another (indicated by plot color). Lower panel: box-plots showing episodic rewards under generalization, for both single-Q (solid boxes) and two-Q (empty boxes) strategies. Using two-Q substantially increases generalization when trained in sparse conditions.}\label{fig:fig4}
 \end{figure}
\subsection{Geometry of recovery}

After a detection, trained Q-learning agents first surge straight upwind; then, they engage in a complex recovery strategy aimed at re-entering the plume, primarily comprising an oscillating forward motion and eventually a downwind return. We find that the relative prevalence of different recovery behaviors depends strongly on plume sparsity. 

To study this phenomenon, we define several metrics which characterize the geometry of a trajectory, as illustrated in Fig.~\ref{fig:fig3}. Additional examples of learned trajectories are shown in the Supplemental Material (see Appendix \ref{app:supp_material}). To wit, these are (1) the upwind distance traversed during the initial surge, (2) the crosswind breadth traversed during casting, (3) the total upwind distance traversed before returning downwind, (4) the total downwind distance traversed during the return, and (5) the time spent searching before the beginning of the downwind return. 

First, we note that 
the upwind surge becomes markedly stronger as the sparsity of the plume increases. This corresponds with the intuition that the surge corresponds to a memory of the last odor detection, and that the typical time between successive detections increases with sparsity.

On the other hand, while the crosswind cast width and downwind return distance appear to increase and decrease, respectively, with plume sparsity, this dependence is quite weak; the strongest signal is that the downwind returns are very short at the highest sparsity level. In Ref.~\cite{heinonen2025optimal}, it was argued that these quantities are connected closely to the wind speed, which was held fixed in our tests.

Finally, in sparse plumes, agents spend more time and cover more distance searching upwind before initiating their downwind return. We reason that this is to avoid becoming trapped far downwind of the source in a region that is particularly poor of  information within sparse plumes. 
On the other hand, in dense plumes, 
the agent begins returning downwind significantly earlier. Drifting downwind will recover the plume when agents accidentally overshoot the source, which is the primary failure mode in dense plumes. %
Indeed, as sparsity increases, agents get lost increasingly further downwind from the source, as shown by the distributions of downwind distances where the last detection occurred, conditioned on the agent failing to find the source (Fig.~\ref{fig:fig3}(d)). Thus, agents fail mainly when they either get trapped far downwind of the plume or overshoot the source; the relative importance of these two failures depends on sparsity of the plume.

By the same logic, agents trained in dense environments generalize poorly to those  sparse environments (Fig.~\ref{fig:fig4}), and vice-versa, as they are unable to execute the appropriate strategy to avoid the failure mode that is more common in the new environment. The poor generalization performance can be attributed to an increase in failure rate (see Supplemental Material in Appendix \ref{app:supp_material}).  As shown in the upper panel of Fig.~\ref{fig:fig4}, agents trained in dense environments fail especially often far downwind of the source when they are tested in sparse environments.

  \begin{figure}
 \includegraphics[width=\linewidth]{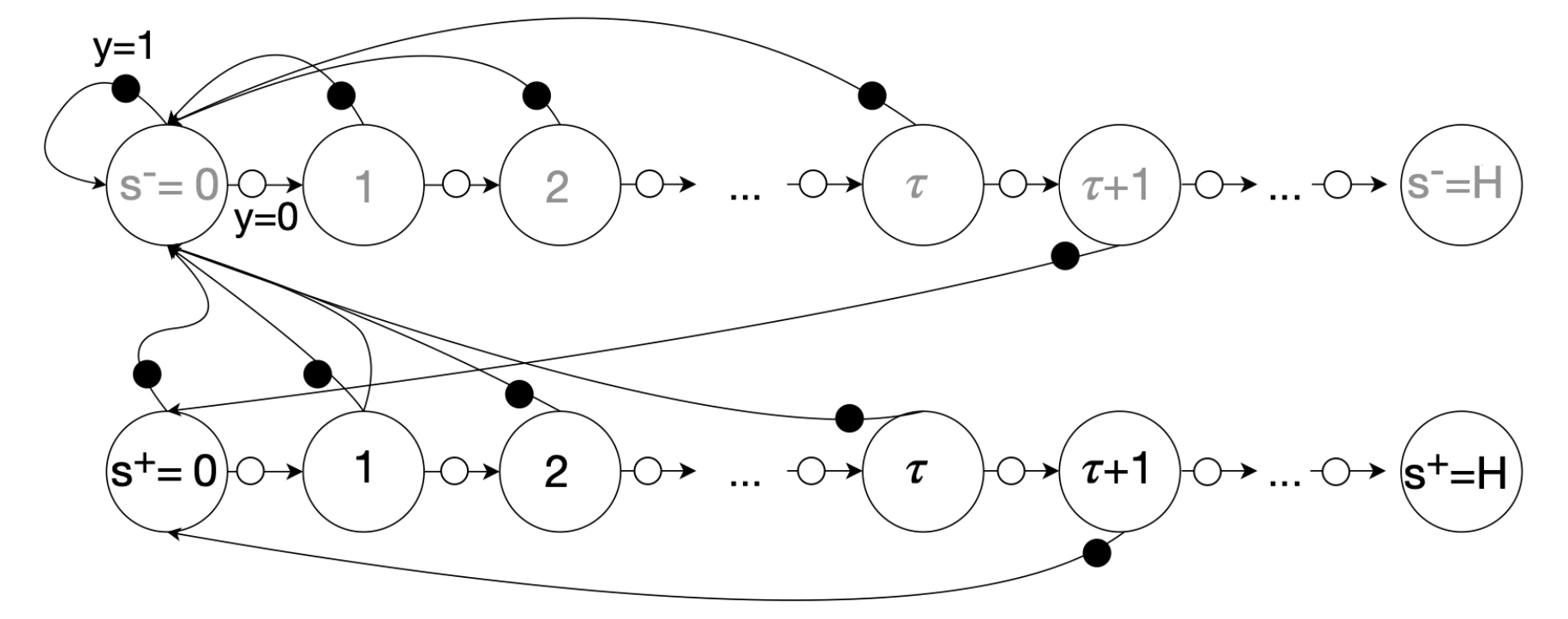}
 \caption{A more flexible algorithm with two separate Q-matrices. The size of the state space has been doubled to $S=S^+ \cup S^- $. Both sets of states are ordered by an integer value, $s^+ \in [0,H]$ and $s^- \in [0,H]$. Odor detections (black circles) reset the agent from any state $s^\pm>\tau$ to state $s^+=0$ and from any state $s^\pm \le\tau$ to $s^-=0$. In the absence of odor detection (white circles), the clock state is incremented: $s^+\rightarrow s^++1$ and ~$s^-\rightarrow s^-+1$. The Q-matrix $Q(s,a)$ is split as $Q^+=Q(s^+,a)$ and $Q^-=Q(s^-,a)$.
}\label{fig:fig5}
 \end{figure}

 \begin{figure*}
 \includegraphics[width=\linewidth]{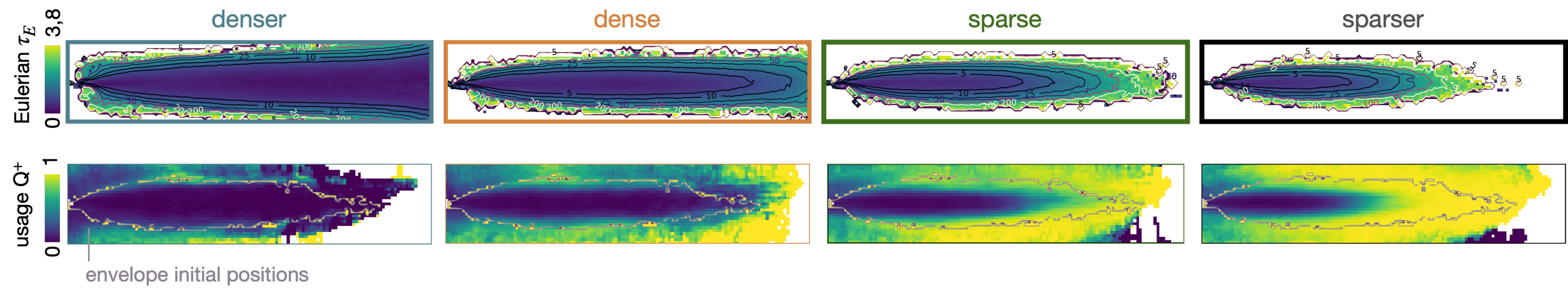}
 \caption{Top: colormap of the average Eulerian blank time (the typical time between successive odor detections) at increasing sparsity level. Contours at $\tau=5,10,25,50,100,200$ are shown. Bottom: usage of $Q^+$, measured as the empirical fraction of test agents using $Q^+$ at each location. $Q^-$ is mostly used within the core of the plume, where detections are frequent, whereas $Q^+$ is mostly used in the back and on the edges, where detections are rare. For the denser plume, $Q^-$ is mostly used, i.e., most agents are in $s\in S^-$.
 } \label{fig:fig6}
 \end{figure*}
 
\subsection{Two-Q agent}
Avoiding overshooting the source and avoiding getting lost in the back of the plume are ultimately conflicting goals, and an agent with a single recovery trajectory cannot simultaneously solve both problems; instead it will simply prioritize whichever problem it is exposed to more often in training. The agent is unable to adapt its behavior to the \emph{local} sparsity level of the plume.

To test if a single agent may learn to solve both problems, we introduce a more flexible algorithm, with two parallel Q-matrices $Q^\pm$ (Fig.~\ref{fig:fig5}). In this setup, the time since the last detection no longer suffices to specify the agent's action; instead, a crude measure of the local sparsity level is also used. In particular, we fix a threshold time $\tau$. Matrix $Q^+$ is employed when the last detection occurred $t>\tau$ after the previous detection%
, and matrix $Q^-$ is employed when it occurred $t\le\tau$ after the previous detection. 

We found that the performance of this two-Q agent did not depend strongly on $\tau$, and a good strategy can be learned and adapted to essentially any threshold (see Supplemental Material in Appendix \ref{app:supp_material}). We chose to select $\tau$ so that both Q matrices are used and thus optimized as close to evenly as possible during training; see the Supplementary Material (Appendix \ref{app:supp_material}) for more details. 

The performance of the two-Q agent is compared to the single-Q agent in Fig.~\ref{fig:fig2}. The benefit of this extra flexibility becomes clear only under sparse conditions, where overshooting and becoming lost far downwind are both important issues to solve; in dense plumes, the latter problem does not occur frequently. For the same reason, using the two-Q strategy also significantly improves the agent's ability to generalize, specifically in the case where the agent was trained in a sparse environment, as shown in Fig.~\ref{fig:fig4} (lower panel).

\begin{figure}
 \includegraphics[width=\linewidth]{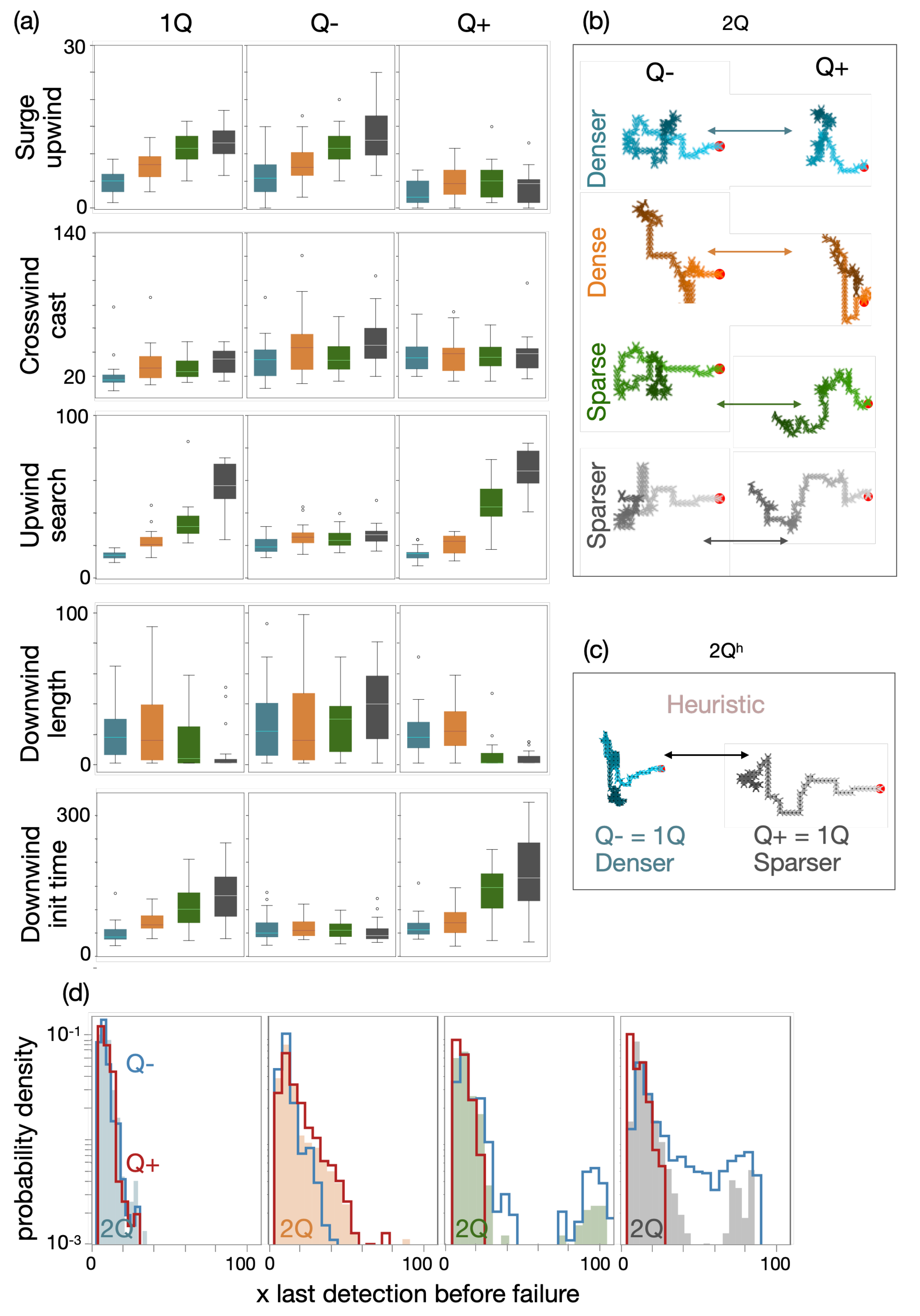}
 \caption{In the two-Q agent, $Q^+$ and $Q^-$ learn distinct recovery strategies
 which target different regions of the plume. 
 (a) Characterization of recovery strategy geometry, as in Figure~\ref{fig:fig3}(c), for $Q^+$ and $Q^-.$ The single-Q result is reproduced for comparison. 
 (b) Sketch of representative policies for both $Q^+$ and $Q^-$ in the four  environments and (c) of the heuristic policy $2Q^h$ (bottom). 
 (d) Downwind positions of the last detection before failure for two-Q, $Q^-$ only, and $Q^+$ only. When using $Q^-$ only, agents fail in the back of the plume as $Q^-$ is not trained to avoid losing the plume there. The reverse happens for $Q^+$, which has not learned to prevent overshooting. 
}\label{fig:fig7}
 \end{figure}

In Fig.~\ref{fig:fig6}, we show that $Q^-$ is more active in the core of the plume close to the source and the centerline, while $Q^+$ is more active in sparse regions at the edges of the plume. Consistent with this observation, we find that $Q^-$ specializes in avoiding overshooting, whereas $Q^+$ specializes in avoiding being trapped at the back of the plume. Indeed, upwind search is far more prevalent in $Q^+$ than in $Q^-$; and it increases with sparsity, similar to trends observed with single $Q$. Conversely, the downwind return occurs much sooner in $Q^-$ (see Fig.~\ref{fig:fig7}(a)). 
$Q^-$ also features a more prominent surge, since it relies on detections more than $Q^+$. Typical trajectories in $Q^-$ and $Q^+$ are compared in Fig.~\ref{fig:fig7}(b). We also show in Fig.~\ref{fig:fig7}(d) the downwind positions of the last detection before failure when using two-Q, only $Q^+$, and only $Q^-$. Using $Q^-$ alone leads to frequent failures in the back of sparse plumes, but sees no significant increase in failure close to the source, again emphasizing its specialization towards downwind returns.
Conversely, using $Q^+$ alone leads to increased failures close to the source, confirming that it does not learn to avoid overshooting.

Finally, since we expect $Q^+$ to be more active in sparse regions and $Q^-$ more active in dense regions, we tested a heuristic policy that uses the Q matrix trained under dense (sparse) conditions as $Q^-$ ($Q^+$) (Fig.~\ref{fig:fig7}). As seen in Fig.~\ref{fig:fig2}, although this policy is fixed and not re-trained in each new environment, it either outperforms or nearly matches the optimized single-Q agent. %

\subsection{A quasi-optimal agent}
The primary shortcoming of the Q-learning approach is its inflexibility; the agent cannot adapt its strategy to its history of odor observations, since it only recalls the time since the most recent detection. To illustrate that an optimal agent does not have this drawback, we trained a quasi-optimal Bayesian agent using the POMDP formalism \cite{kaelbling1998}. This approach retains memory of its observation history in the form of a spatial map, called the belief $b$, assigning probability to possible source locations, which is updated using a model for the detection probability. This form of memory is lossless if the model is perfect (in fact, we have neglected that observations are correlated in time \cite{heinonen2025exploring}). The agent's policy maps beliefs to actions, so that in principle the entire observation history may influence its behavior.

\begin{figure}[h!]
\centering
 \includegraphics[width=0.65\linewidth]{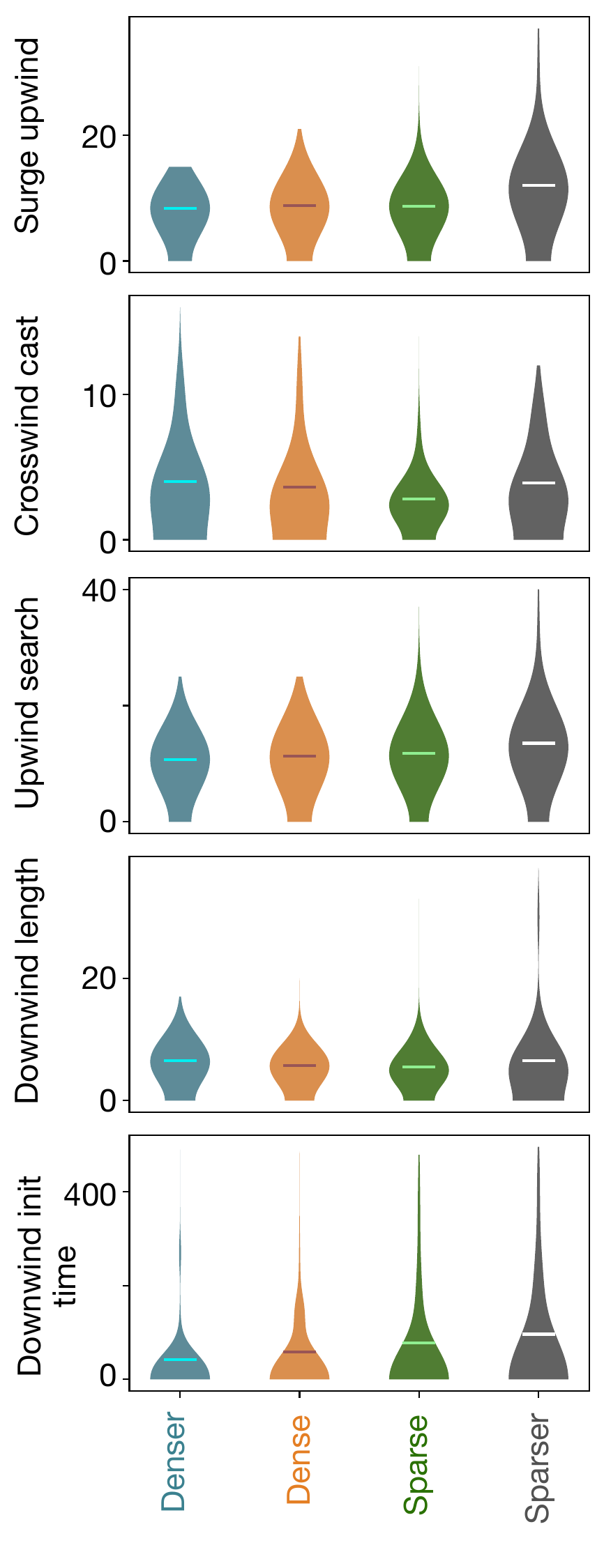}
 \caption{Geometry of a single quasi-optimal Bayesian %
recovery strategy varies from trial to trial. 
Distributions of the five trajectory metrics are shown as violin plots; these are obtained by tracking a single agent's behavior during a long sequence of odor blanks, for an ensemble of prior observation histories. The metrics can take broadly different values depending on the history, reflecting the flexibility granted by the Bayesian agent's sophisticated, high-dimensional memory. %
Consistent with %
the trends observed across Q-agents, the distributions of upwind surge length, upwind search length, and downwind initiation time exhibit increasingly long tails as sparsity increases.}\label{fig:fig8}
 \end{figure}

As shown in Fig.~\ref{fig:fig8}, a single agent's trajectory after losing the plume may indeed exhibit significantly different geometry depending on its detection history. However, certain trends persist; for one, the typical surge length and time until downwind return continue to increase with sparsity, which are realized primarily as lengthenings in the tails of the distributions of those quantities. This maximum flexibility affords greater performance (Fig.~\ref{fig:fig2}), but at the cost that vastly greater computational resources are required to execute the policy.

\section{Discussion}\label{sec:discussion}
Despite the notorious difficulty of the olfactory search problem in turbulent settings, we have shown that a simple discrete memory specifying the time since the last detection suffices to develop a high-performing agent. The learned policy specifies only how to recover from loss of the plume, underlining that deciding what to do in the absence of odors is the most difficult and most important part of the problem, as previously noted (see e.g.~Refs.~\cite{heinonen2025optimal,rando2025,verano2023}). The present work also shows that the classical cast-and-surge heuristic, which shares the same set of internal states, can be improved significantly by optimizing the precise shape of the casting and surging and additionally learning to return downwind.

We have also shown how plume sparsity influences the geometry of recovery strategies. In dense plumes, returning downwind is essential to avoid overshooting the source; while this is also important under sparse conditions, sparse plumes also require more significant upwind motion to avoid trapping at the sparse rear of the plume. The inability to reconcile both of these behaviors is a key weakness of the single-Q approach, which can be partially mitigated by extending the internal memory in a minimal way, as demonstrated with the two-Q approach. A Bayesian agent, on the other hand, is capable of quasi-optimal performance by adapting its strategy to any arbitrary sequence of odor observations; in addition to much larger computational resources, executing such a policy requires an accurate model of the environment.

It should be noted that how exactly the agent negotiates the tradeoffs discussed here depends to some extent on the initial condition. 
Our agents are optimized for efficient search starting from a fixed area that is sampled uniformly and kept identical in all environments.
If instead the search starts from a region that adapts to sparsity of the plume, or if a detection is forced to initiate the search, trends are qualitatively preserved while details vary (see Supplementary information in Appendix \ref{app:supp_material}). 
The range of initial positions must be appropriately considered, as it can vary for a variety of reasons connected to prior information both for animals~\cite{carde2021_annurev} and in robotic applications~\cite{ishida2006mobile}.

Here, we labeled actions relative to the mean wind, which is supposed to be known in advance by the agents. Measuring the mean wind is however unrealistic because it requires long integration which may be impractical in robotics and unrealistic in biology. 
In reality, animals rely on local measures of flow speed,  using e.g.~antennas for insects~\cite{mean_flow,Bell_Kramer79,Suveretal2019,Okuboetal2020}, whiskers for rodents~\cite{whiskers} or the lateral line for marine organisms~\cite{lateral_line}. 
Forthcoming research will study anemometric agents that align their actions relative to a shifting wind axis which they estimate in real time~\cite{piro2026}. 

Our results also suggest a concrete way to test how much plume recovery is governed by a fixed internal program \emph{vs} adapts to recent olfactory history. Note that a history dependence in navigating insects has been observed, including variable upwind surge lengths, but the origin of this variation is poorly understood~\cite{carde2021_annurev,kuenen1994,van_bruegel_delay,pang2018history}.
In the clock-state agent, once odor is lost, the recovery trajectory is determined by a single scalar variable, the time elapsed since the last detection, producing a single stereotyped sequence comprising surge, lateral exploration, and eventual downwind return. By contrast, the Bayesian agent shows that even under the same immediate condition of odor loss, different prior detection histories can produce markedly different recovery trajectories. 
This distinction suggests an experimental protocol: condition the searcher on different recent odor histories, then impose an identical prolonged blank, and compare the ensuing recovery trajectories. In animals, such tests could be implemented in virtual-odor (i.e., optogenetic) experiments, similar to, for example, Ref.~\cite{jayaram2023}. In robots, our results suggest that controllers with highly compressed memory may already be sufficient for robust plume tracking, depending on the acceptable tolerance for failure probability. 

\begin{acknowledgments}
This research was supported by grants to AS from the European Research Council under the European Union’s Horizon 2020 research and innovation programme (grant agreement number 101002724 RIDING), the National Institutes of Health under award number R01DC018789 and by a France 2030 support managed by the Agence Nationale de la Recherche, under the reference ANR-23-PEIA-0004 (PDE-AI project). This work represents only the views of the authors; the European Research Council Executive Agency and the other funding
agencies are not responsible for any use that may be made
of the information it contains. 
\end{acknowledgments}

\section{Methods}\label{sec:methods}

\subsection*{Data description}

Data used to train and test the agents are set of $2598$ matrices $\{D_t\}^{2598}_{t=0}$ where every matrix
$D_t \in \mathbb{R}^{123 \times 27}$ contains the odor intensity in every position $(i, j)$ i.e. $(D_t)_{i,j}$ represents the odor intensity in position $(i, j)$ at timestep $t \in \mathbb{N}$. The odor source is located at position $x_\text{source} =[0, 13]$; positions in a circle of radius $2$ centered at the source ({\it source region}) are considered terminal states. The odor dataset is a downsampled version of odor simulation used in \cite[Simulation 1]{rando2025} where spatial dimensions are reduced by a factor of $10$. Data information are summarized in Table \ref{tab:data_info}. The four environments (Denser, Dense, Sparse, Sparser) are obtained by thresholding the data, setting all values below the {\it noise threshold} $n_\text{thr} \in \mathbb{R}_+$ to zero. In particular, for the Denser environment $n_\text{thr} = 10^{-6}$, for Dense $n_\text{thr} = 3 \times 10^{-6}$, for Sparse $n_\text{thr} = 5 \times 10^{-6}$, and for Sparser $n_\text{thr} = 7 \times 10^{-6}$. 
The dataset and code will be shared shortly.

\begin{table}[H]
    \centering
    \caption{Summary of dataset details}
    \label{tab:data_info}
    \begin{tabular}{ll}
        \toprule
        Feature &  Value \\
        \midrule
         \# Grid points & $123 \times 27$\\
         \# Time slices & $2598$\\
         Agent speed (grid point/ time slice) & $1$\\
         Source position & $[0, 13]$\\
         Source region radius & $2$\\
         \bottomrule
    \end{tabular}

\end{table}

\subsection*{Agent design and state representation}

In this section, we describe the internal state representations used by the single-$Q$ and two-$Q$ agents, and how these states are updated over time.

\paragraph*{Single-$Q$ agent.}
For a single-$Q$ agent, the internal state is a clock variable $s \in \mathbb{N}$ that counts the number of consecutive time steps during which the observed signal remains below a fixed sensitivity threshold $c_{\mathrm{thr}} > 0$. At each time step $t \in \mathbb{N}$, the agent receives an observation $\omega_t \in \mathbb{R}_+$ and updates its clock accordingly. Specifically, if $\omega_t \ge c_{\mathrm{thr}}$, the agent detects odor and the clock is reset to $s_t = 0$. If $\omega_t < c_{\mathrm{thr}}$, no odor is detected and the agent is considered to be in a void state; in that case, the clock is incremented according to $s_t = s_{t-1}+1$.

At time step $t=0$, the agent receives an initial observation $\omega_0 \in \mathbb{R}_+$ and initializes its clock as
\begin{equation*}
s_0 =
\begin{cases}
0, & \text{if } \omega_0 \ge c_{\mathrm{thr}},\\
1, & \text{otherwise}.
\end{cases}
\end{equation*}
For each subsequent time step $t>0$, the clock is updated as
\begin{equation*}
s_t =
\begin{cases}
0, & \text{if } \omega_t \ge c_{\mathrm{thr}},\\
s_{t-1}+1, & \text{otherwise}.
\end{cases}
\end{equation*}

In both training and testing, agents are allowed to execute at most $H \in \mathbb{N}_+$ steps (the \emph{horizon}). This implies that this class of agents has $H$ states, with state $0$ corresponding to odor detection and the remaining $H-1$ states corresponding to void states.

\paragraph*{Two-$Q$ agent.}
For a two-$Q$ agent, the internal state is again based on a clock, but the agent additionally records the duration of the previous odor-free interval. More precisely, at each time step $t \in \mathbb{N}$, the agent stores the length of the most recent completed blank interval in a variable $\tau_t$, and compares it with a fixed threshold $\tau$.

Unlike the single-$Q$ agent, the two-$Q$ agent uses two distinct families of void states. State $s_t=0$ again indicates that odor has just been detected, i.e.\ $\omega_t \ge c_{\mathrm{thr}}$. When a void observation is encountered, i.e.\ $\omega_t < c_{\mathrm{thr}}$, the subsequent evolution depends on whether the previous blank interval was shorter or longer than $\tau$. If $\tau_t \ge \tau$, the agent uses one family of void states, indexed by $1,\dots,H-1$; otherwise it uses a second family, indexed by $H,\dots,2H-1$. In total, the two-$Q$ agent therefore uses $2(H-1)+1$ states.

If an odor detection occurs, the variable $\tau_t$ is updated to the length of the blank interval that has just ended, namely $s_{t-1}$, and the clock is reset to $s_t=0$. If instead a void observation occurs, then the update depends on the previous clock value. If $s_{t-1}=0$, the new state is set to either $1$ or $H$, depending on whether $\tau_t < \tau$ or $\tau_t \ge \tau$. If $s_{t-1}>0$, the clock is simply incremented.

We initialize the previous blank length as $\tau_0=\tau+1$, and define the initial clock state by
\begin{equation*}
s_0 =
\begin{cases}
0, & \text{if } \omega_0 \ge c_{\mathrm{thr}},\\
H, & \text{otherwise}.
\end{cases}
\end{equation*}
For each $t>0$, the previous blank length is updated as
\begin{equation*}
\tau_t =
\begin{cases}
s_{t-1}, & \text{if } \omega_t \ge c_{\mathrm{thr}},\\
\tau_{t-1}, & \text{otherwise},
\end{cases}
\end{equation*}
and the clock state is then updated according to
\begin{equation*}
s_t =
\begin{cases}
0, & \text{if } \omega_t \ge c_{\mathrm{thr}},\\
H, & \text{if } \omega_t < c_{\mathrm{thr}} \,\wedge\, s_{t-1}=0 \,\wedge\, \tau_t \ge \tau,\\
1, & \text{if } \omega_t < c_{\mathrm{thr}} \,\wedge\, s_{t-1}=0 \,\wedge\, \tau_t < \tau,\\
s_{t-1}+1, & \text{if } \omega_t < c_{\mathrm{thr}} \,\wedge\, s_{t-1}>0.
\end{cases}
\end{equation*}

At initialization, $\tau_0$ is set above the threshold $\tau$. This choice makes the agent initially behave as though it were in a sparse region far from the source, thereby triggering the corresponding exploratory strategy. If the agent instead begins near the source, where blank intervals are typically shorter, the value of $\tau_t$ adjusts rapidly after a few detections.

\paragraph*{Characterization of trajectories in the void.}
To characterize behavior in the absence of odor, we analyze the deterministic trajectory generated by an agent when it starts at position $(0,0)$ and thereafter receives only blank observations. From this trajectory we extract five geometric features: surge length, upwind search, backtracking length, cast width, and backtracking initiation time.

These quantities are defined as follows:
\begin{enumerate}
\item \textbf{Surge length.} This is the number of initial upwind steps (leftward moves) taken by the agent before it leaves the surge phase. To avoid counting isolated deviations caused by optimization noise, we use a tolerance of two steps: the surge is deemed to end only after two consecutive non-leftward actions. %
\item \textbf{Cast width.} This quantifies the lateral extent of the largest cast and is approximated by the difference between the maximum and minimum $y$ positions reached by the agent.
\item \textbf{Upwind search.} This measures the extent of the trajectory in the upwind direction and is computed as the minimum $x$ position attained by the agent.
\item \textbf{Backtracking length.} This is the extent of the trajectory in the downwind direction and is approximated by the absolute difference between the largest and smallest $x$ values reached.
\item \textbf{Backtracking initiation time.} This is the number of steps executed before the beginning of sustained downwind motion. Operationally, we define it as the number of actions taken before the agent performs three consecutive rightward moves.
\end{enumerate}

The values of the upwind search, backtracking length, and cast width depend on how much of the void trajectory is included. Because the final part of the trajectory may be noisy---the deepest void states may have been visited only rarely during training---we evaluated these metrics under several truncation criteria: (i) the full trajectory; 
(ii) the largest void state visited during the last $500$ training episodes (the criterion used in the main text); 
(iii) the average void state visited during the last $500$ training episodes; and 
(iv) the maximum duration of a blank during the last $500$ training episodes; and
(v) the mean duration of a blank during the last $500$ training episodes (a blank goes from one detection to the next detection, so it excludes instances where the agent gets lost). All choices lead to qualitatively similar patterns (see Supplemental Material \ref{app:supp_material}).

\subsection*{Agent behavior and policies}

We now describe how the agents interact with the environment to solve the olfactory navigation problem. At each time step $t \in \mathbb{N}$, the agent receives an odor observation $\omega_t \in \mathbb{R}_+$, updates its internal state $s_t$, and chooses an action $\bm{a}_t$. The action space consists of the four cardinal directions:
\begin{equation*}
\mathcal{A} := \{\bm{e}_1,\bm{e}_2,-\bm{e}_1,-\bm{e}_2\},
\end{equation*}
where $\bm{e}_i \in \mathbb{R}^2$ denotes the $i$th canonical basis vector.

Agents are allowed to leave the data grid; once outside the grid, they continue to receive zero observations. If the agent reaches the source region, \textcolor{black}{it receives a distinguished observation $\omega_t=-1$. Formally, let $\bm{x}_\text{source} \in \mathbb{R}^2$ denote location of the odor source. The source region $S$ is defined as follows.}
\begin{equation*}
S = \{\bm{x} : \|\bm{x} - \bm{x}_\text{source} \|_2 \le 2\}.
\end{equation*}
\textcolor{black}{In our experiments, the odor source is $\bm{x}_\text{source} = [0, 13]$.}

At each time step, the agent selects an action using an $\varepsilon$-greedy tabular Q-learning policy. Let $Q$ denote the agent's Q-matrix. Then
\begin{equation*}
\bm{a}_t =
\begin{cases}
\bm{a} \in \arg\max_{\bm{a}\in\mathcal{A}} Q(s_t,\bm{a}), & \text{with probability } 1-\varepsilon,\\
\bm{a} \sim \mathcal{U}(\mathcal{A}), & \text{with probability } \varepsilon,
\end{cases}
\end{equation*}
where $\mathcal{U}(\mathcal{A})$ denotes the uniform distribution on $\mathcal{A}$. During training, $\varepsilon$ is gradually decreased across episodes. During testing, we set $\varepsilon=0$, so that the policy is deterministic.

\subsection*{Training procedure}

Each episode begins at a random location and a random time slice of the DNS data. The initial position is sampled uniformly from the set of grid points at which there is a nonzero probability of detecting odor in the sparsest environment, i.e.\ from the set of points for which the concentration exceeds $7\times 10^{-6}$ at least once. The same initialization rule is used for all threshold values, to ensure consistency across environments.

At each time step $t$, the agent receives an observation $\omega_t \in \mathbb{R}_+$, updates its state $s_t$, and selects and executes an action $\bm{a}_t$. After the action is executed, the agent receives a reward $r_t$ and a new observation $\omega_{t+1} \in \mathbb{R}_+$. The reward is defined by the position reached after the action: if the new position lies outside the source region, then
\begin{equation*}
r_t = -(1-\gamma),    
\end{equation*}
where $\gamma \in (0,1)$ is the discount factor; if the new position lies inside the source region, then $r_t=1$.

The initial Q-matrix is set to $Q_0(s,\bm{a})=-1$ for all states and actions. For each episode $k$, the Q-matrix is updated at every step $t=0,\dots,H$ according to
\begin{widetext}
\begin{equation*}
Q_{k+1}(s_t,\bm{a}_t)
=
(1-\alpha_k)\,Q_k(s_t,\bm{a}_t)
+
\alpha_k\left(
r_t + \gamma \max_{\bm{a}\in\mathcal{A}} Q_k(s_{t+1},\bm{a})
\right).
\end{equation*}
\end{widetext}
Here $s_t$ is the state at time step $t$, $\bm{a}_t$ is the action selected by the $\varepsilon_k$-greedy policy, and $s_{t+1}$ is the state obtained from the new observation $\omega_{t+1}$.

The learning rate is scheduled as
\begin{equation*}
\alpha_k
=
\alpha_{\mathrm{end}}
+
(\alpha_{\mathrm{init}}-\alpha_{\mathrm{end}})
\max\left\{
\frac{\alpha_{\mathrm{decay}}-k-1}{\alpha_{\mathrm{decay}}},\,0
\right\},
\end{equation*}
with $\alpha_{\mathrm{init}}=0.1$, $\alpha_{\mathrm{end}}=10^{-4}$, and $\alpha_{\mathrm{decay}}=2.5\times 10^6$.

Similarly, the exploration parameter is scheduled as
\begin{equation*}
\varepsilon_k
=
\varepsilon_{\mathrm{end}}
+
(\varepsilon_{\mathrm{init}}-\varepsilon_{\mathrm{end}})
\max\left\{
\frac{\varepsilon_{\mathrm{decay}}-k-1}{\varepsilon_{\mathrm{decay}}},\,0
\right\},
\end{equation*}
with $\varepsilon_{\mathrm{init}}=1.0$, $\varepsilon_{\mathrm{end}}=10^{-4}$, and $\varepsilon_{\mathrm{decay}}=2.5\times 10^6$.

Thus, at episode $k=0$ we have $\alpha_0=0.1$ and $\varepsilon_0=1.0$, while after $2.5\times 10^6$ episodes both parameters remain fixed at $10^{-4}$. {\color{black} With this choice, we strike a balance between speed and accuracy. Note that tabular Q-learning is only guaranteed to converge asymptotically anyways. Additionally, there is no guarantee that the optimal policy is deterministic under partially observability \cite{jaakkola1994}.} The horizon is fixed at $H=500$ steps, the discount factor is $\gamma=0.998$, and training is run for $3{,}000{,}001$ episodes.

The sensitivity threshold $c_{\mathrm{thr}}$ is set to $10^{-6}$ for the \emph{denser} plume, $3\times 10^{-6}$ for the \emph{dense} plume, $5\times 10^{-6}$ for the \emph{sparse} plume, and $7\times 10^{-6}$ for the \emph{sparser} plume. For each setting, we train $20$ single-$Q$ agents and $20$ two-$Q$ agents using the parameters above.

\subsection*{Evaluation}

To evaluate performance, we consider three metrics: the cumulative reward $G$, the normalized arrival time, and the success fraction $f^+$. These quantities are averaged over all valid initial positions.

Valid initial positions are the points $(x,y)\in D_{\mathrm{init}}$, where $D_{\mathrm{init}}$ is the set of grid points outside the source region for which there is a nonzero probability of observing an odor concentration greater than $7\times 10^{-6}$, corresponding to the sparsest environment. For $(x,y)\in D_{\mathrm{init}}$, let $T(x,y)$ be the number of steps required for the policy to reach the source, and let $T_{\min}(x,y)$ be the shortest possible arrival time from $(x,y)$.

We define the normalized arrival time as
\[
\frac{T_{\min}(x,y)}{T(x,y)}
\]
if the trajectory reaches the source region, and as $0$ otherwise. The success fraction $f^+$ is the fraction of valid initial positions from which the policy reaches the source. Finally, $G$ denotes the cumulative reward averaged over all valid initial positions.

\subsection*{Bayesian POMDP agent}

To implement and solve the POMDP, we follow the procedure used previously in \cite{loisy2022,loisy2023,heinonen2023,heinonen2025optimal,heinonen2025exploring}. The agent's partially observed state is its position $\bm{x}$ relative to the source, evolving on a two-dimensional grid of size $(2M-1)\times(2N-1)$, where $M\times N$ are the dimensions of the simulation box.

The observation space consists of three symbols: \emph{whiff}, \emph{blank}, and \emph{source}. The likelihoods $\Pr(\mathrm{whiff}\mid \bm{x})$ and $\Pr(\mathrm{blank}\mid \bm{x})$ are given by the empirical probabilities that the concentration at $\bm{x}$ exceeds or falls below, respectively, the threshold $c_{\mathrm{thr}}$. The observation \emph{source} occurs if and only if the agent is in the source region, which is treated as a single absorbing state. Because the observation model is assumed known, the Bayesian POMDP approach is model-based, whereas Q-learning is model-free.

The agent maintains a belief state $b_t(\bm{x})$, i.e.\ a posterior distribution over positions. With a slight abuse of notation, we treat the source region as a single state. At each time step, the agent receives an observation $\omega_t$, selects an action $\bm{a}_t\in\mathcal{A}$, and updates its physical state deterministically via $\bm{x}\mapsto \bm{x}+\bm{a}_t$. The belief is then updated by Bayes' rule:
\begin{equation*}
b_{t+1}(\bm{x}\mid \omega_t,\bm{a}_t)
=
\frac{
b_t(\bm{x}-\bm{a}_t)\Pr(\omega_t\mid \bm{x}-\bm{a}_t)
}{
\sum_{\bm{x}'} b_t(\bm{x}'-\bm{a}_t)\Pr(\omega_t\mid \bm{x}'-\bm{a}_t)
}.
\end{equation*}

The initial belief is uniform on the initialization region $D_{\mathrm{init}}$. The agent receives reward $1$, discounted at rate $\gamma_{\mathrm{POMDP}}$, upon entering the source region, and $0$ otherwise. Thus,
\[
R(\bm{x},\bm{a}) =
\begin{cases}
1, & \text{if } \bm{x}+\bm{a}\in S,\\
0, & \text{otherwise}.
\end{cases}
\]

The optimal expected return can be expressed as a functional of the current belief $b$, namely the value function $V[b]$. This function satisfies the Bellman optimality equation \cite{kaelbling1998}
\begin{widetext}
\begin{equation}\label{eq:bellman}
V[b_t]
=
\max_{\bm{a}\in\mathcal{A}}
\left\{
\sum_{\bm{x}} b_t(\bm{x})R(\bm{x},\bm{a})
+
\gamma_{\mathrm{POMDP}}
\sum_{\omega}
\Pr(\omega\mid b_t,\bm{a})\,
V\!\left[b_{t+1}(\cdot\mid \omega,\bm{a})\right]
\right\}.
\end{equation}
\end{widetext}
The unique solution specifies the optimal policy by selecting, at each belief state, an action that maximizes the right-hand side.

Because this Bellman equation is computationally intractable except through approximation, we use the SARSOP algorithm \cite{kurniawati2008} to compute a quasi-optimal policy for each value of $c_{\mathrm{thr}}$. In these computations we use $\gamma_{\mathrm{POMDP}}=0.98$ and terminate SARSOP after a fixed runtime of $4000$ seconds.

To obtain an ensemble of recovery trajectories for the quasi-optimal agent, we use a constrained Monte Carlo tree procedure. The agent executes a search trial in the DNS data, and each time it makes a \emph{whiff} observation we create a branch point. Along one branch, the search continues normally; along the other, all subsequent observations are forcibly set to \emph{blank} for $H=500$ time steps. The trajectory generated during this forced-blank sequence is recorded.

To downweight highly implausible forced-blank continuations, each such trajectory is assigned the weight
\[
w
=
\prod_{t=t_{\mathrm{off}}}^{T_{\mathrm{back}}}
\Pr(\mathrm{blank}\mid \bm{x}_t),
\]
where $t_{\mathrm{off}}$ is the time at which the forced-blank sequence begins, and $T_{\mathrm{back}}$ is the time at which the agent completes its downwind return.

Each search trial is terminated when the agent reaches the source region or when the initial search time reaches $100$ steps. We repeat this procedure $100$ times for each threshold value.

\subsection*{Cast-and-surge heuristic}

As a baseline, we implemented a variant of the cast-and-surge heuristic of Balkovsky and Shraiman \cite{balkovsky2002}, adapted to a discrete gridworld and augmented with a free parameter controlling the opening angle of the search cone. The algorithm is given in Algorithm~\ref{alg:cast_surge_angle}. The upwind distance traveled between successive casts was fixed at one grid unit. For each sparsity level, we tuned the half-opening angle in increments of $5^\circ$ so as to maximize the episodic reward $G$.

\begin{algorithm}[H]
\caption{Angle-parameterized cast-and-surge policy on a grid}
\label{alg:cast_surge_angle}
\begin{algorithmic}[1]
\State \textbf{Input:} half-opening angle $\theta \in (0,\pi/2)$ and observation $\omega \in \{0,1\}$
\State \textbf{Internal state:} $x$ (upwind distance since last detection), $y$ (lateral displacement), $s \in \{-1,+1\}$ (casting side)
\State $\sigma \gets \tan\theta$
\If{$\omega = 1$}
    \State $(x,y,s) \gets (1,0,+1)$
    \State \Return $(-1,0)$ \Comment{surge upwind}
\EndIf
\State $y^\star \gets s\lceil \sigma x \rceil$
\If{$y < y^\star$}
    \State $y \gets y+1$
    \State \Return $(0,1)$
\ElsIf{$y > y^\star$}
    \State $y \gets y-1$
    \State \Return $(0,-1)$
\Else
    \State $s \gets -s$
    \State $x \gets x+1$
    \State \Return $(-1,0)$
\EndIf
\end{algorithmic}
\end{algorithm}

\bibliographystyle{plain}
\bibliography{references}
\newpage

\onecolumngrid
\appendix 

\section{Supplementary Material}\label{app:supp_material}

\begin{figure}[H]
\centering
 \includegraphics[width=0.9\linewidth]{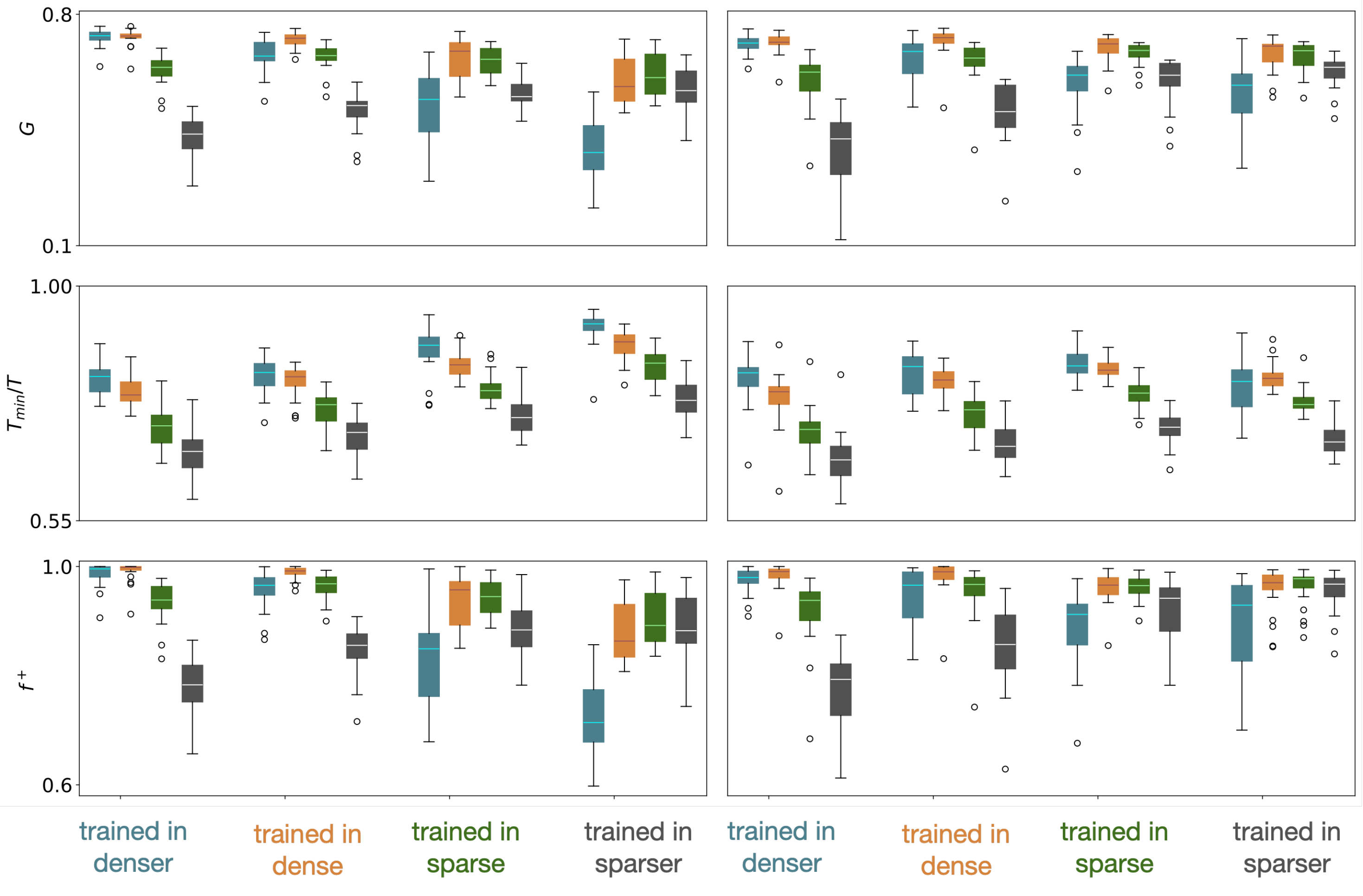}
 \caption{Generalization of single Q agents (left) and two-Qs agents (right), measured by the cumulative reward $G$ (top tow) and its projections on the normalized time to reach the target (mid row) and the fraction of successes (bottom row). Training in sparse plumes provides particularly visible benefits for success rates in generalization.} 
 \end{figure}

 \begin{figure}[H]
\centering
 \includegraphics[width=0.9\linewidth]{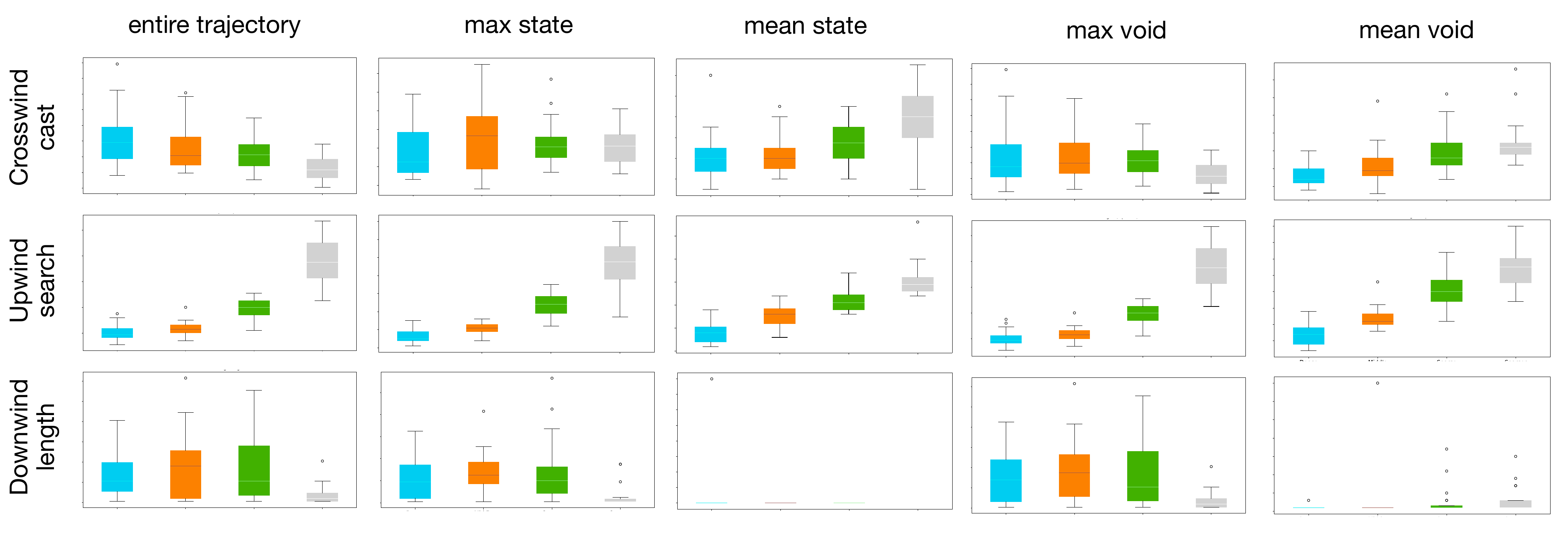}
 \caption{Cast width, upwind search and downwind length, defined in the main text, depend on the duration of the trajectory that is considered. Left to right: results using truncation criteria described in Materials and Methods, \emph{Characterization of trajectories in the void}, (i) to (v). Second column: results with criterion used throughout the main text. %
 }
\end{figure}

 \begin{figure}[H]
 \centering
 \includegraphics[width=0.9\linewidth]{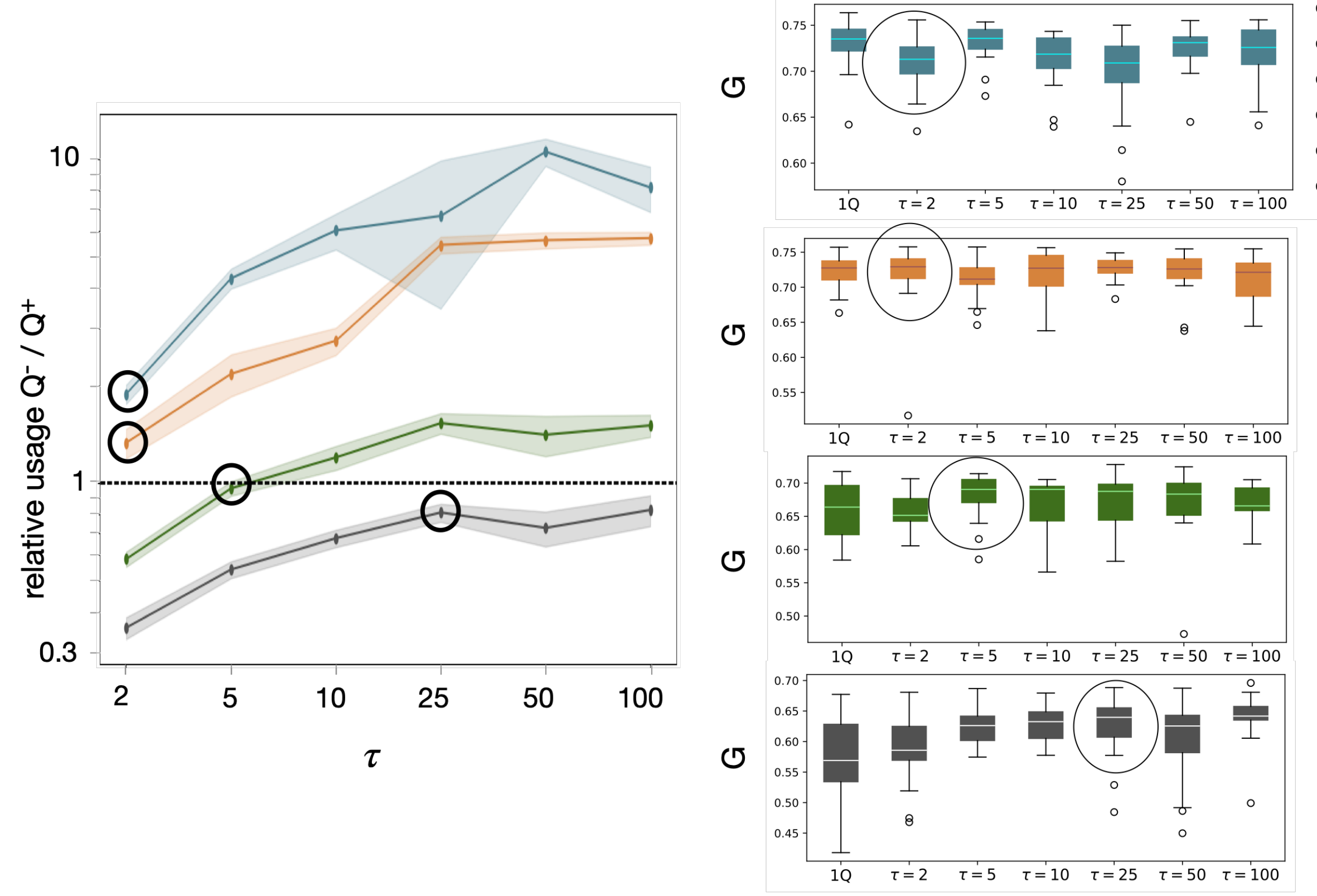}
 \caption{Results of the 2Q algorithms are robust to the choice of the threshold $\tau$ that dictates which Q-matrix is used (Right). We choose $\tau$ so that both Q-matrices are used as evenly as possible (left). }
 \end{figure}

 \begin{figure}[H]
 \centering
 \includegraphics[width=0.9\linewidth]{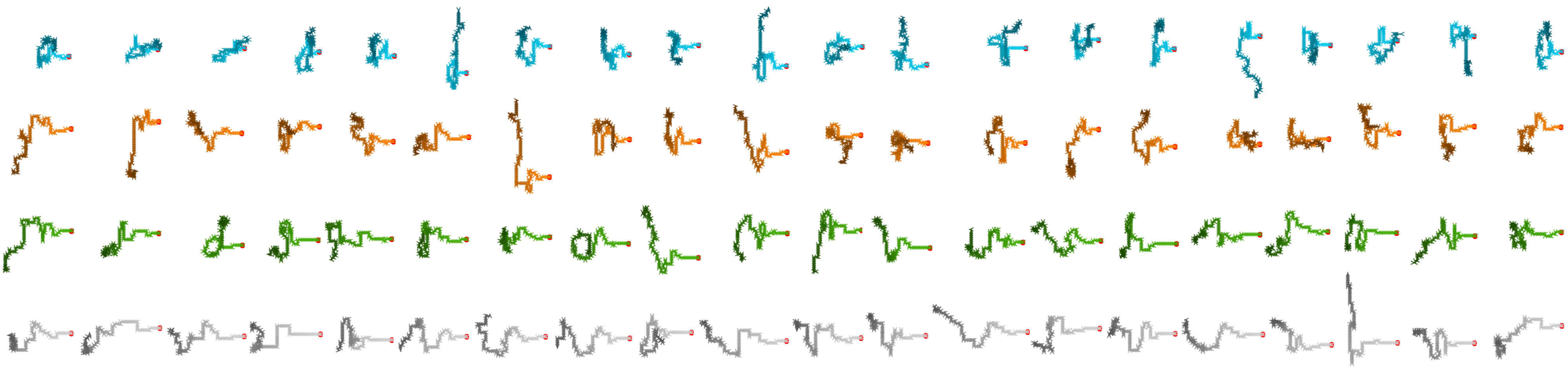}
 \caption{Policies of the 20 single Q agents, trained in the four environments with increasing sparsity from top to bottom. The red dot marks the position for state $s=0$, successive steps indicate policy for increasing values of the state variable $s>0$, with darker and darker colors}
 \end{figure}

 \begin{figure}[H]
 \centering
 \includegraphics[width=0.9\linewidth]{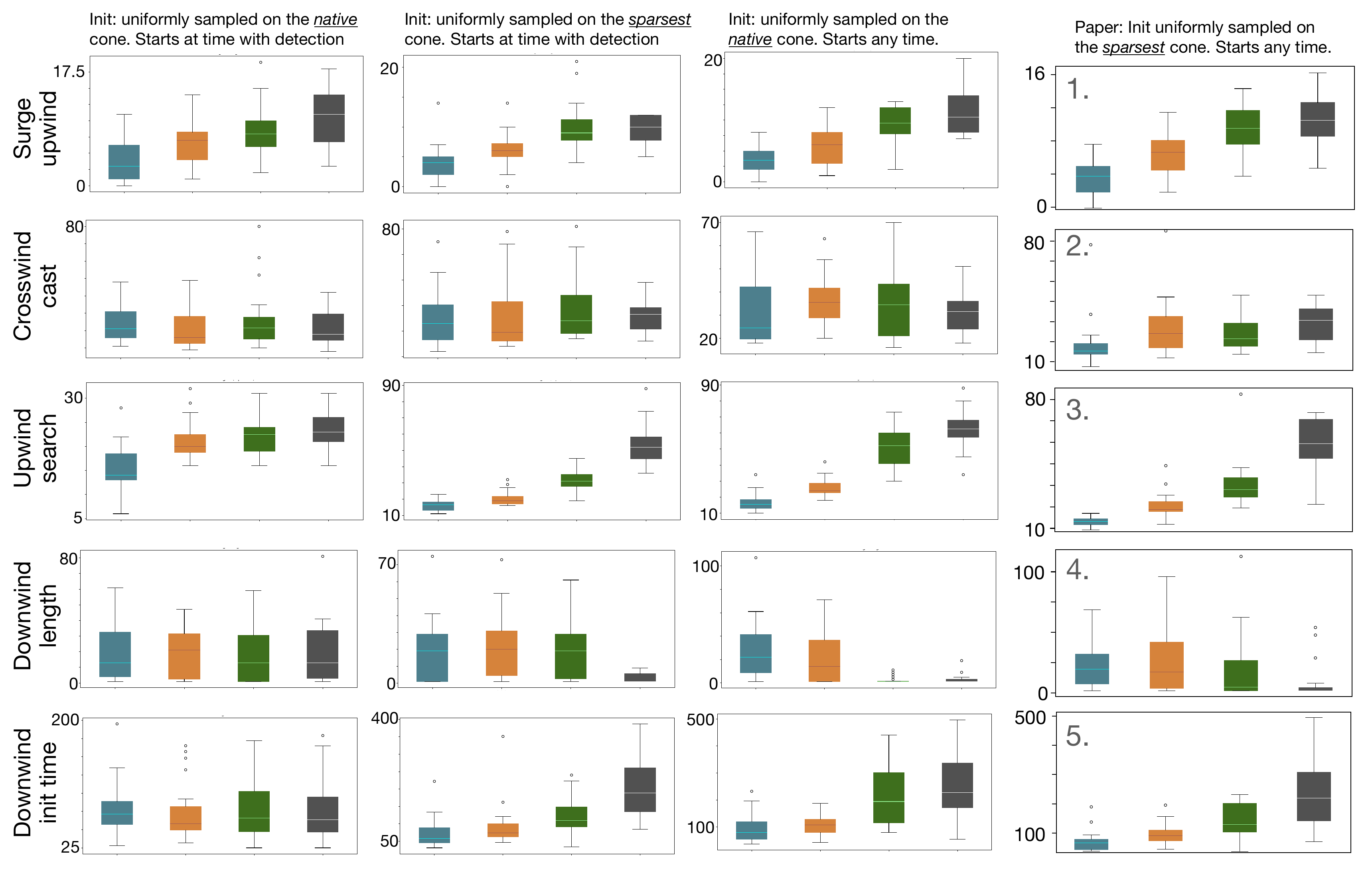}
 \caption{Supplementary. Optimizing agents starting with different initial conditions yield qualitatively similar patterns in geometry of trajectory in the void. First two columns from left:
 Agents start at a timestamp within the simulation when they detect odor; Last two columns on the right: agents start at a random timestamp within the simulation, when they may or may not detect. In all cases, agents start at a location that is sampled uniformly from a region that encompasses all points where the odor is above threshold at any time. First and third column from left: the threshold is the same that defines the training environment, thus increases with sparsity. Note that for this choice, denser plumes are optimized to reach the target from a region that is larger than dense plumes.
 Second and fourth column: threshold is fixed at the highest value  so that the initial condition is the same for all training environments. Rightmost column is used throughout main paper. All patterns are robust except for downwind initiation and length for initial condition that changes with the environment and forcing a detection at time 0.}
 \end{figure}

\end{document}